# Small MMI with simplified coherent coupling branches used to develop a lower footprint power splitter on SOI platform

R. PEYTON[1,2], J. MARTINEZ VALDIVIEZO,[1,2] D. PRESTI[1,2] AND

G. A. TORCHIA[1,2,*]

*[1]Centro de investigaciones Ópticas—CONICET-CIC-UNLP, Camino Centenario y 506 s/n, M.B.Gonnet (1897), Buenos Aires. Argentina.*

*[2]Departamento de Ciencia y Tecnología, Universidad Nacional de Quilmes, Bernal (1876), Buenos Aires, Argentina.*

** gustavot@ciop.unlp.edu.ar*

**Abstract:** In this paper we introduce a new approach to fabricate under the Silicon on Insulator (SOI) platform a 1x2 power splitter and scale it to 1xN system device from the proposed model. The current strategy of design is based on the well-known simplified coherent coupling (SCC) which allows the construction of very sharp bends. To split the incoming light, we add at the entrance of the splitter a small Multi Mode Interferometer (MMI). The sensibility of the design parameters is analysed and discussed along the paper. In addition, we fabricated and evaluated the new splitter architecture on 1x2 and 1x4 versions and then compared our devices with those splitters based on standard multimode interferometer (MMI) also performed in the same SOI photonic chip. The new splitter scheme shows a similar performance than standard MMI, however a slight spectral flat response was observed in new devices. By the novel approach, highly compact devices can be designed so this strategy can also open new avenues to improve and enhance the performance of integrated devices developed under the SOI scheme.

## 1. Introduction

Currently, silicon photonic technology has a significant importance in optical communications systems since its small footprint devices, CMOS process compatibility and high integration level. In this contribution, we focus on silicon-on-insulator (SOI) technology which is the most popular platform for that purpose [1]. The high refractive index contrast between silicon and silica layers makes successful submicron waveguides and support tight bends. In addition, the state-of-the-art electronics processes developed in actual industry foundries can be easily exploited. However, unlike electronic circuits where electrical routing can be accomplished flexibly, optical routing is limited either for optical conditions or technological limitation. Each design must be performed considering the design rules of a CMOS process. For example, almost fundamental components (e.g. Y-branch) used in integrated photonics show better performance if sharp corners and spline interpolation are conducted, but usually they can violate the design rules describing fabrication process details [2]. In order to solve the issues, geometries based on simplified coherent coupling, multi-sectional bends and other concepts can be explored.

Simplified coherent coupling (SCC) is a technique to bend light through small straight waveguides with sharp bending. Some studies have already reported on bent waveguides and splitters with low-loss, compact and efficient devices compared with other types of geometries. Physically this effect relies on when decoupled mode light in a curve can be efficiently coupled back into a subsequent guiding path if the phase difference between the modes (guided and unguided) is an odd multiple of π. Consequently, from this theory, the bending loss results in an oscillatory function. This depends on the section length for each splitter branch as well as on the difference between the effective index of the guided mode and the weighted average effective index corresponding to all unguided modes. Finally, the optical transmission also noticeably relies on the propagation wavelength due to coupling between guided and radiated modes [3-6]. In addition, in this paper, our power splitter includes a small MMI at the entrance which was strategically located in order to split the input mode, so each splitting mode travels along each branch path of the splitter and then propagates along the curves by means of the simplified coherent coupling theory.

Regarding to the power splitter designs and its description extracted from the literature, we can point out the main strategies used to develop this optical system as follows: 1) ordinary Y-branch architecture with s-line for the branches paths, 2) Multimode Interferometer (MMI) based power divider and 3) plasmonic power splitters. The designs 1 and 2 are the most used currently in different technological applications such as: optics communications, optical sensors, and these schemes also presents a strong impact regarding to new photonic equipment and instruments. The key point in both designs relies on the way to fabricate that: on one side, impact to the balanced output power from the exit branches [7-11]. This issue is associated to the fabrication tolerance from photolithography and associated techniques used to perform these kinds of devices. On the other side, the actual uncertainties on the final dimensions of these photonics devices due to fabrication process represents the main weakness for this technology. However, this drawback can be overcome considering sizeable optical circuits since the use of larger photonic structures with same fabrication tolerance (below 5%) give rise to low loss devices.

Finally, the group of power splitters designed using plasmonic schemes which are subwavelength systems present a well-known large energy loss due to thermal effects produced by electrons moving at resonance plasmons. However, in the last year, there exist a lot of hybrid devices by using the plasmonic concepts to achieve a power divider in combination with silicon guiding structures. This new scheme results in a more efficient integrated system to transmit light through the whole tiny new device getting insertion losses like pure silicon photonic structures [12-14].

From last year, a new platform using silicon on insulator substrates by adding organic materials has been used to produce photonics integrated devices. This scheme is so-called as Silicon–organic hybrid (SOH) photonics platform [15]. This complementary material drives to the silicon integrated photonics devices electro-optics properties which is very necessary for signal modulation in optical communications as well as for control of tuneable wavelength filters based on ring resonators. Therefore, the organic material thin films would represent a suitable system for fitting these requirements so in addition many other interesting properties are contributed such as low power consumption for fabrication, recyclability, and self-healing among other characteristics. Following this development line, Lithium Niobate thin films were also integrated with the silicon platform, so this new hybrid system fits the requirements for optical modulation at the hundreds of GHz range which is nowadays strongly demanded by data transmission systems [16,17].

In this paper, we propose an innovative design of power splitter instead of those based on the well-known Y-branch and/or MMI structures. Our proposed device combines the simplified coherent coupling theory (SCC) to construct the splitter branches with high aperture angles. At the entrance, a small multimode interference (MMI) is used to split the input light in two exit modes which then are guided on each branch. We present a device with a high transmission efficiency and a small footprint. The photonic component was designed on silicon-on-insulator platforms. In this paper, computational simulations were performed to study the relationship between the transmitted power and the set of design parameters. Moreover, we establish a free design parameter that can be adjusted according to the desired application, and thus achieve a nicely adaptable device.

In addition, in this work we also fabricate our design with the modern CMOS photonics process and then compare the performance of our splitter design with those standard MMI splitters provided by the manufacturer, which were integrated into the same photonic chip developed in a SOI wafer.

## 2. Materials and Methods

The Silicon on Insulator (SOI) 1x2 and 1x4 power splitters were fabricated on a 340-nm-thick crystalline silicon core on top of a buried oxide layer with a thickness of 2 μm on a silicon substrate and a silicon dioxide deposition of 1 μm as final cladding. Standard single mode strip waveguide width for λ=1.55 μm results 450 nm. The UV lithography technique and complementary methods were conducted to define the photonics circuits, following the standard procedure as it is well described in reference [8,18].

In addition, to carry out a complete optical characterization, testing and quantifying the performance to the optical splitter devices included in the photonic chip a deep analysis was conducted. To do that a 6-axis freedom degree motorized station was used. This system allows sub-micrometres movement for the X, Y and Z and the Euler angles can be set by this equipment. To couple the laser light at communication band was made by using of grating coupling scheme. In this sense, the strategy to optimize the coupling was conducted by using fibre array and follow a specific procedure as describe: the SOI chip was mounted in a X, Y motorized stages and the fibre array was located on a platform allowing total angular and z movements in order to be easy the light coupling process.

This motorized station is controlled by software working in a PC and can evaluate and optimize the best position to each device into the photonics chip with respect to the fibre array by performing an initial point software. To assess the power splitter, to couple and collect the in/out signals we used a V-grove 1x8 fibres array from OZ-Optics company (Canada). Additionally, a 15 mW tuneable laser from Shenzhen Box Optronics Technology (China) working from 1530 nm to 1560 nm (C-band) stepping by 50 GHz (approx. 0.4 nm) was conducted to analyse the optical transmission response for the splitters studied in this work. Finally, by using a headed germanium power meter connected to the PM101U - Power Meter Interface from Thorlabs Inc. (USA) the output optical signal was registered. To assist the coupling process several CCD cameras are located to carry out a detailed top and lateral view for the system chip-fibre array.

Complementary, the computational simulations were made by the 3D finite-difference time-domain (3D-FDTD) methods [19]. The design draw of the power splitters presented in this work is sketched in Figure 1. From this picture, it can be seen the optical length paths used to construct the integrated device based on the simplified coherent coupling (SCC) are clearly detailed. Additionally, at the entrance a small multi-mode interference (MMI) device is used to divide the incoming light. All the design parameters to be optimized for this new approach are: L1, L2, and θ which correspond to the length of path 1 and path 2 for the branches, and the aperture angle of the splitter, respectively. From this picture is also point out that the path L2 which corresponds to the region where the radiation modes are uncoupling/coupling to the L1 paths within each branch of the splitter.

In addition, this picture shows LMMI and WMMI which correspond to the length and width, respectively for MMI located at the entrance of the splitter. All these details are also depicted in Figure 1. As it can be expected, from SOI platform a small footprint device can be reached since the effective bent of the proposed design is more abrupt than a typical bend waveguide commonly used for several power splitter strategies [9-12].

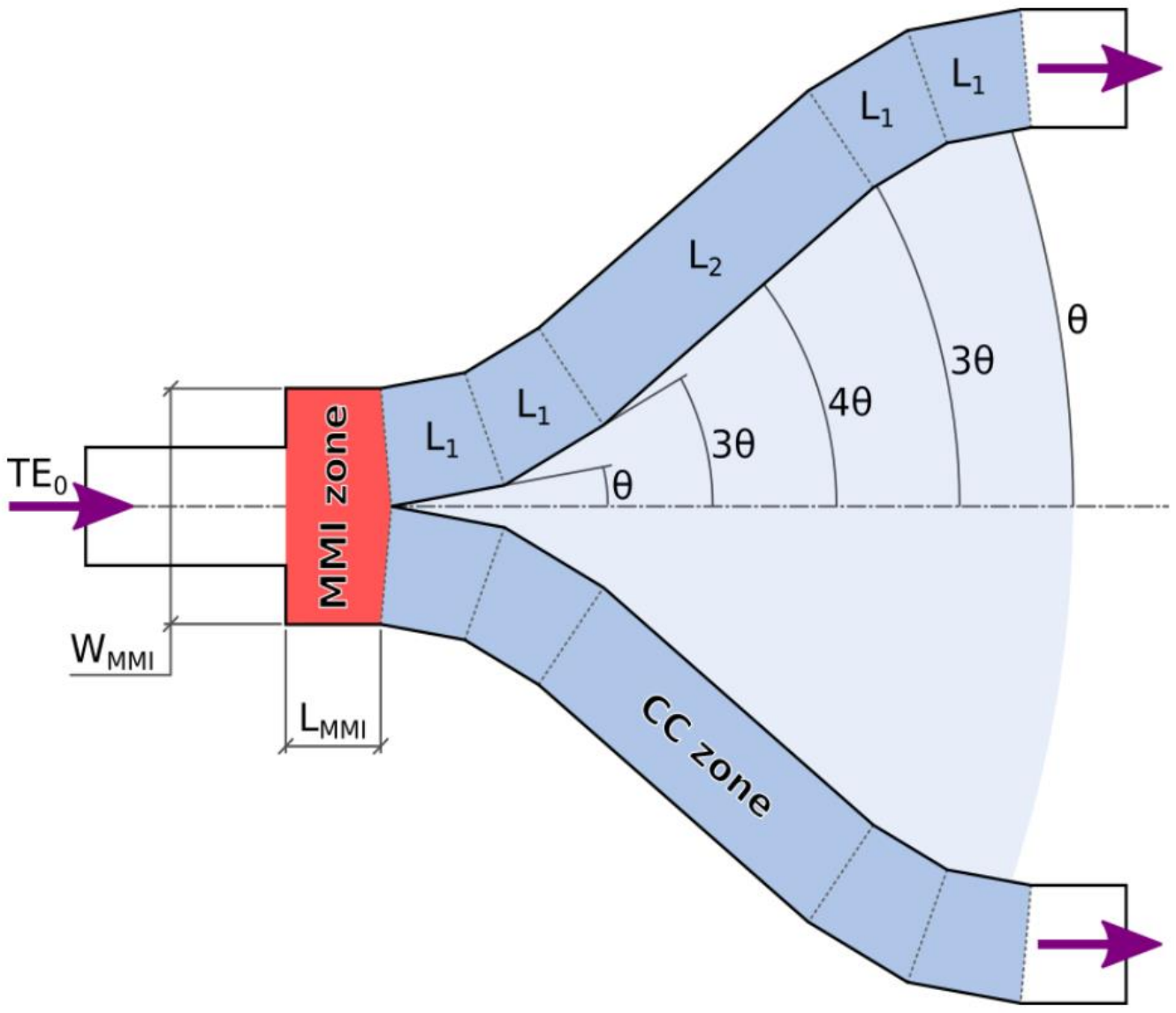

Fig. 1. Schematic draw corresponding to the proposed power splitter based on a small MMI at the entrance and simplified coherent coupling theory (SCC) for the output branches. A mode TE0 is coupled to the designed power splitter (SCC). All design parameters are also highlighted in this picture.

In Figure 2 (a) we present the ***gds*** layout made for the fabrication of photonic chips that includes several photonics designs and types of power splitter analysed in this work. In particular, the proposed design that is constructed by a small MMI following with the output branches follow the SCC architecture made in several version, 1x2, 1x4 and 1x8. As it can be seen from this picture, every device is between a reference optical circuit, that is a 1x1 input/output circuit. This design is used to test the coupling losses to the chip, and this can be used to disengage this value for the experimental measurements taked from the splitters. The distance between each one of the couplers is 127 μm, this value was used for all other devices manufactured. This distance corresponds to a standard separation of arrays of optical fibres to couple and decoupled light from every photonic device. As it can be seen in this part of photonic chip many circuits were achieved so we pay attention in this work to the power splitter based on SCC and MMI design are detailed in the Figure 2 panel b) and c).

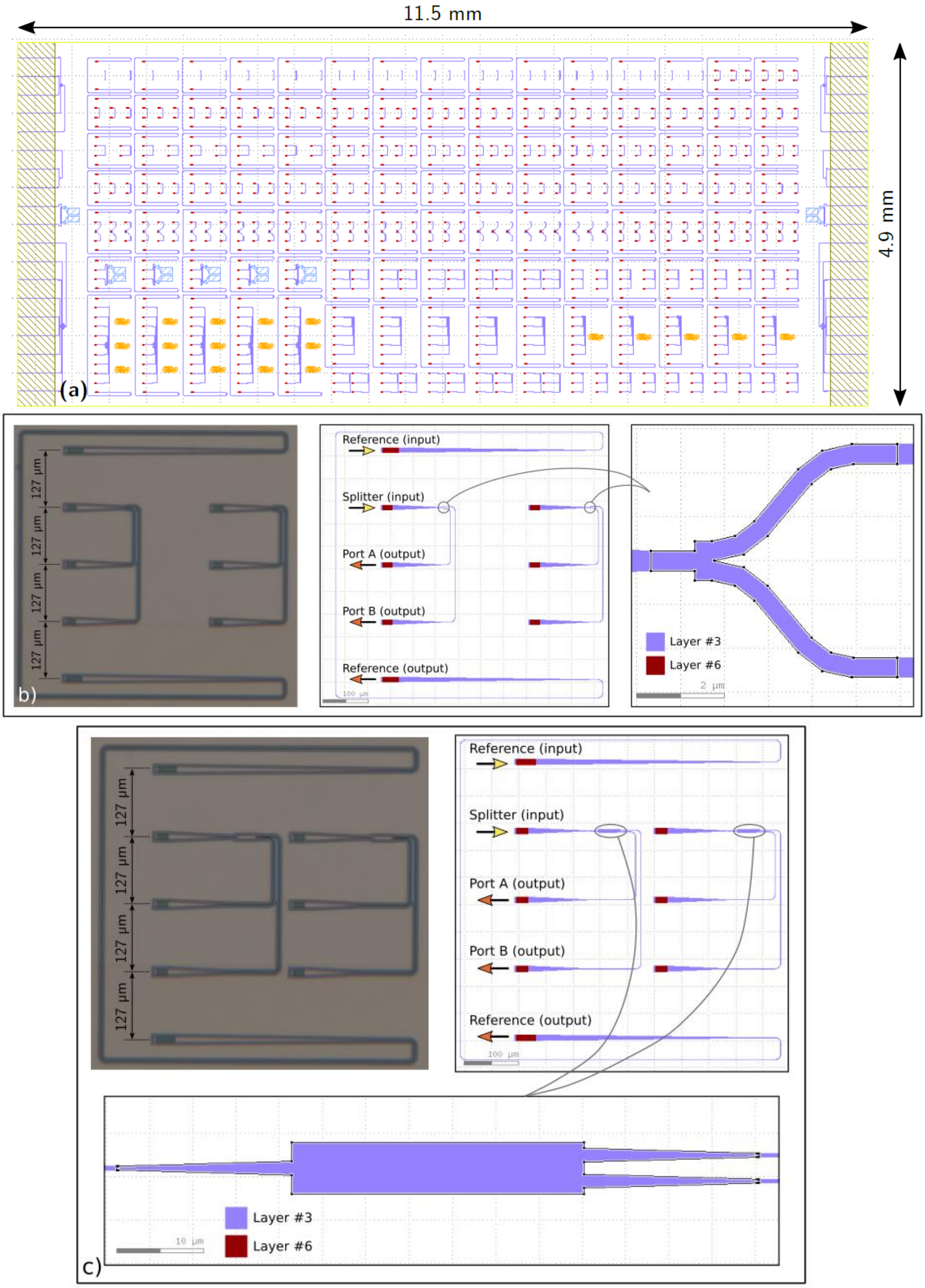

Figure 2 (a) Layout corresponding to the photonic chip used in this work. This includes the power splitters 1x2 and 1x4 designed and tested in this paper (top). Panel (b) shows details for SCC splitter design and a schematic draw for this system (on the right). Panel (c) presents the standard MMI splitter layout on our photonic chip (on the top) and describes details of this splitter design (on the bottom).

In Figure 2 panel (b) is presented the proposed splitter design, in particular the 1x2 version is sketched. From this picture is appreciable the mentioned SCC splitter architecture, it can observe a small MMI at the entrance to split the propagation modes and after that we can see the two high deflection branches which were designed follow the Simplified Coupling Coherent (SCC) theory for each output paths (on the right). In addition, on the middle of this panel is detailed the 1x1 optical circuit to couple input/output signals which was used as a power signal reference to decouple the optical losses by conducting the grating coupling procedure to the chip.

Finally, standard 1x2 power splitter based on multimode interference (MMI) is shown on the bottom of the Figure 2. Principally, the idea to include this device is to carry out a comparative characterization of propagation performance with our design. The current MMI design is provided by the manufacturer itself under a standard design and is often used to create more complex photonic circuits. In this sense, we will compare the performance of the two types of 1x2 power splitter under coupling experiments performed at the C communication band. As you can see, as a first inspection, the footprint of our design is significantly smaller than the standard MMI device provided by the manufacturer.

## 3- Results and discussion

### *3.1. Computational Simulations*

From the simplified coherent coupling (SCC) theory, the most important parameter to be determined is the length of L1. This parameter will define the oscillatory (periodic) behaviour of the structure, so we have studied the dependence of the power transmission versus the length of L1 path. To simplify the analysis, the simulation has been made by assuming L2 = L1 and $\theta = 10^{o}$, note that it is useful to write L2 in terms of L1 times. As shown in Figure 3, there are many maximum and minimum peaks of transmission. This peak structure is a kind of periodic function as a function of path length L1. In addition, we found that the higher transmission is reached at the first period of this curve. This fact is getting when L1 is equal to 0.4 μm. From this first result as the goal of this kind of structure is achieved as a small footprint device, we take the shortest length in the first period of transmission.

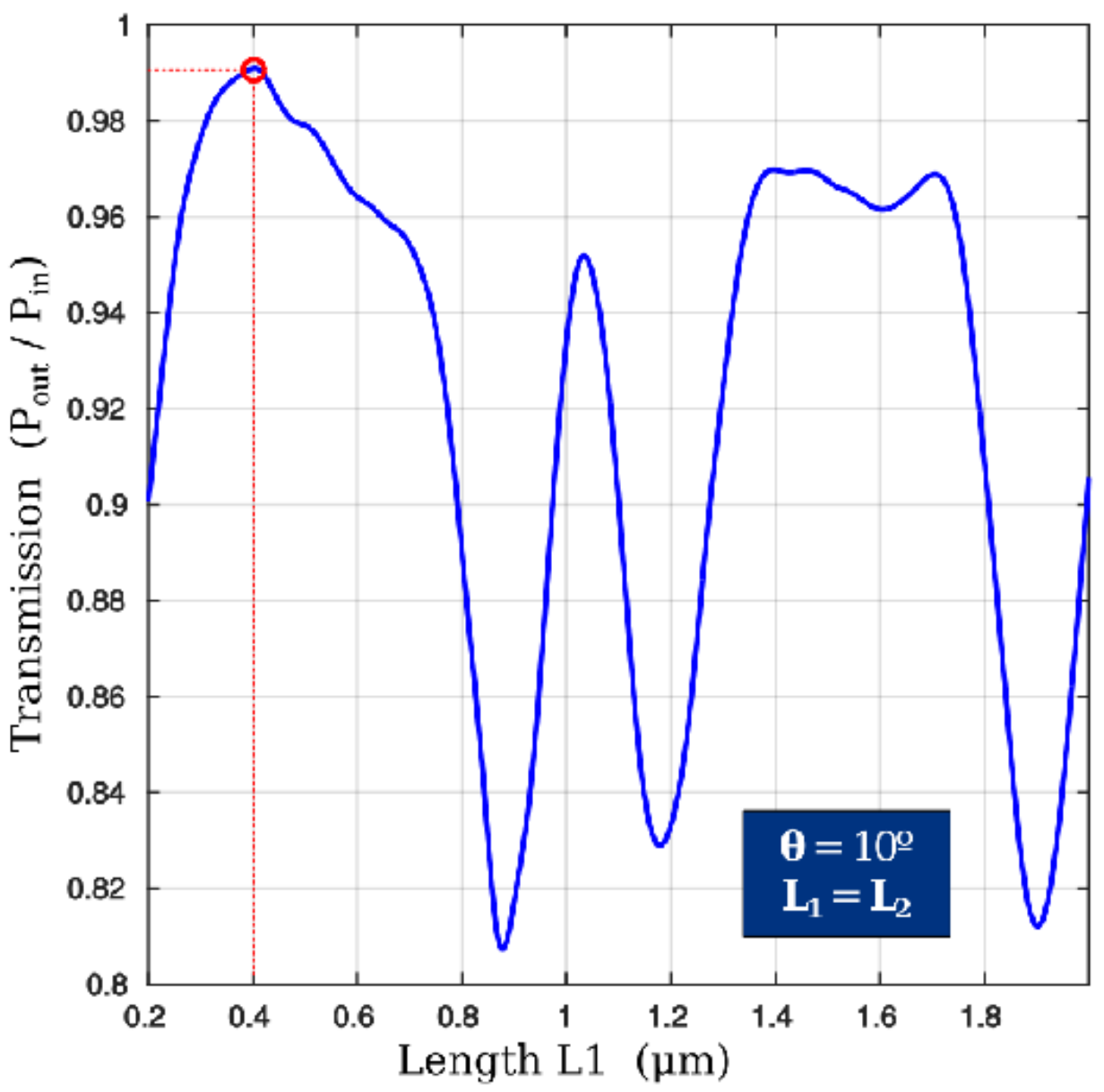

Figure. 3. Computational results corresponding to the optical transmission for the SSC splitter explored close to the coherent coupling zone by sweeping the length L1. It was considered L2 = L1 and θ = 10 °.

To study the sensibility of the design parameters previously fixed, we must analyse the change in the optical transmission with respect to L2 and θ variations. For this purpose, we have implemented two simulations which are shown in Fig. 4. The power transmission by varying L1 and L2 at the same time when θ equal to 10° is displayed in Figure 4(a). In the same sense, Figure 4(b) shows the transmission by simultaneously sweeping L1 and θ, when L2 is equal to L1. White dashed lines in both figures represent the first period of transmission, the same that is plotted in Figure 2. If we pay attention to the difference of the minimum transmission between Figure 4(a) and Figure 4(b), we can see that the design is clearly more sensitive to fluctuation of the angle θ than the length of the path L2. This is due to a large angle means a significant increase of the energy transported by unguided modes. As a result, only a small portion of that energy is coupled back into the waveguide, while the rest is lost as radiated energy to the cladding [3, 4]. On the other hand, the simulation results also indicate the path L2 is a parameter strictly related to the simplified coherent coupling, and hence a periodic behaviour of the transmission is observed. According to Figure 2 and Figure 4(a), the first maximum (0.4 μm) is less sensitive than the second one (1.15 μm), so it is convenient to use that first maximum to reach a robust design. Therefore, we adopt L1 = 0.4 μm and θ = 12.5° whereas the path length L2 will be a free parameter to conveniently adjust. This parameter can be selected either to reduce the excess propagated loss or to improve the wavelength response.

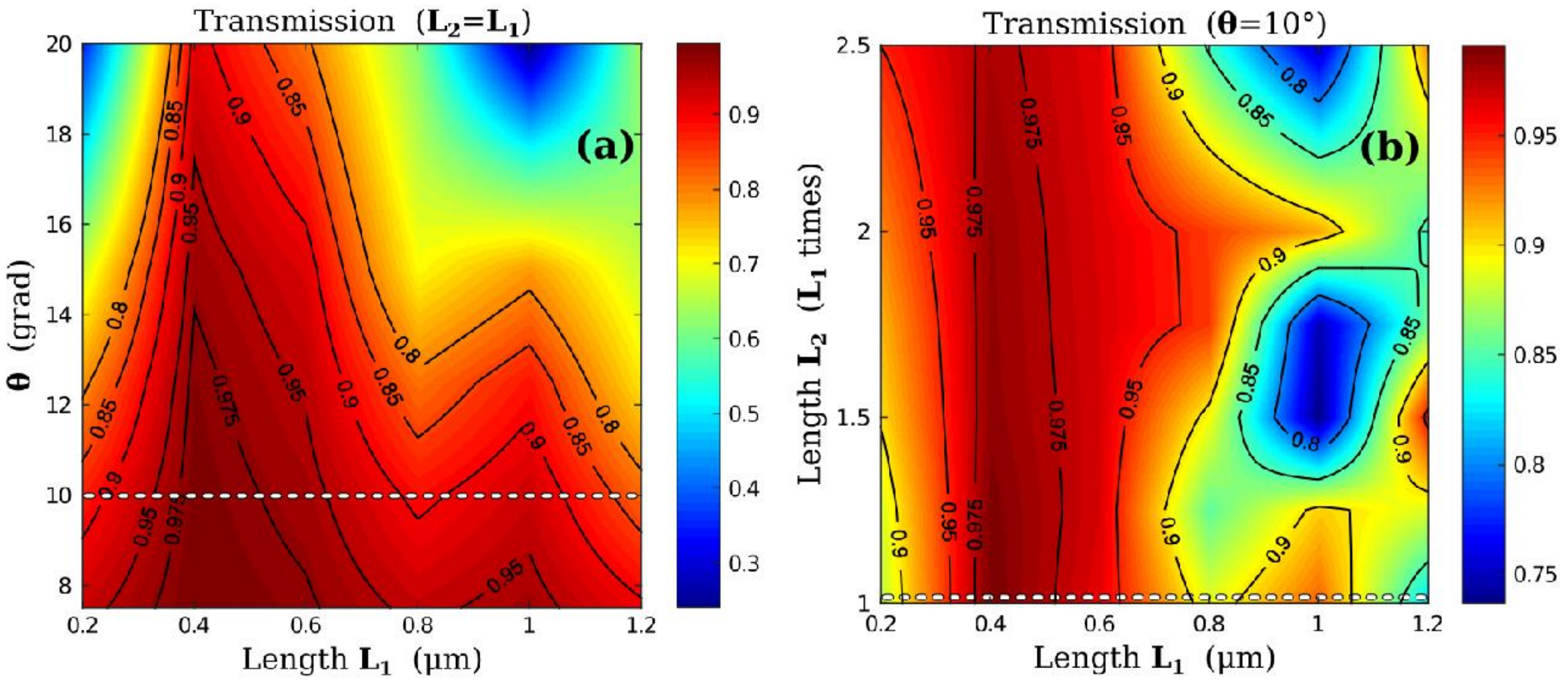

Figure 4. Optical transmission sensitivity within a coherent coupling period as a function of L1 with respect to: panel (a) design angle (θ) and on panel (b) of length (L2).

As commented before, a multimode interference design is proposed to divide the light into two branches. Figure 1 shows the schematic configuration of the MMI waveguide suggested. This, based on self-imaging effect, is a widely used component in the integrated photonics which can perform many different splitting and combining functions [20-22]. Even though the key structure of an MMI device is a waveguide designed to support high excited modes, we used a small enough structure that supports only three TE modes. From a constructive perspective, it is convenient to use a multimode interference waveguide with a small width, in our case it is twice of the single-mode waveguide. As the characteristic self-imaging length LMMI has a quadratic dependence on the width WMMI, the footprint will remain very small if we use that size of multimode structure [23]. For all that, we choose WMMI = 0.9 μm in our device. We initially simulate the propagation of the light by considering an infinite length MMI structure; the interference pattern for this situation is shown in Figure 5(a). The input electric field is reproduced at a specific length with slight phase retardation, and the modes in this MMI region also become only symmetrical modes. One should keep in mind that the wavelength dependence in many applications cannot be in fact, it was found to have lower wavelength sensitivity than structures with larger WMMI dimensions [24]. Once the propagation has been studied, we proceed to define the optimal length in which two images of the incoming single mode are reproduced. Simulations were performed by setting the free parameter L2 at 0.4 μm and sweeping the LMMI length. The results of these simulations are illustrated in Figure 5(b), where the power transmission in one of the two branches is plotted as a function of the multimode interference length. As we can see in this figure, the best condition of splitting was found at LMMI = 0.415 μm.

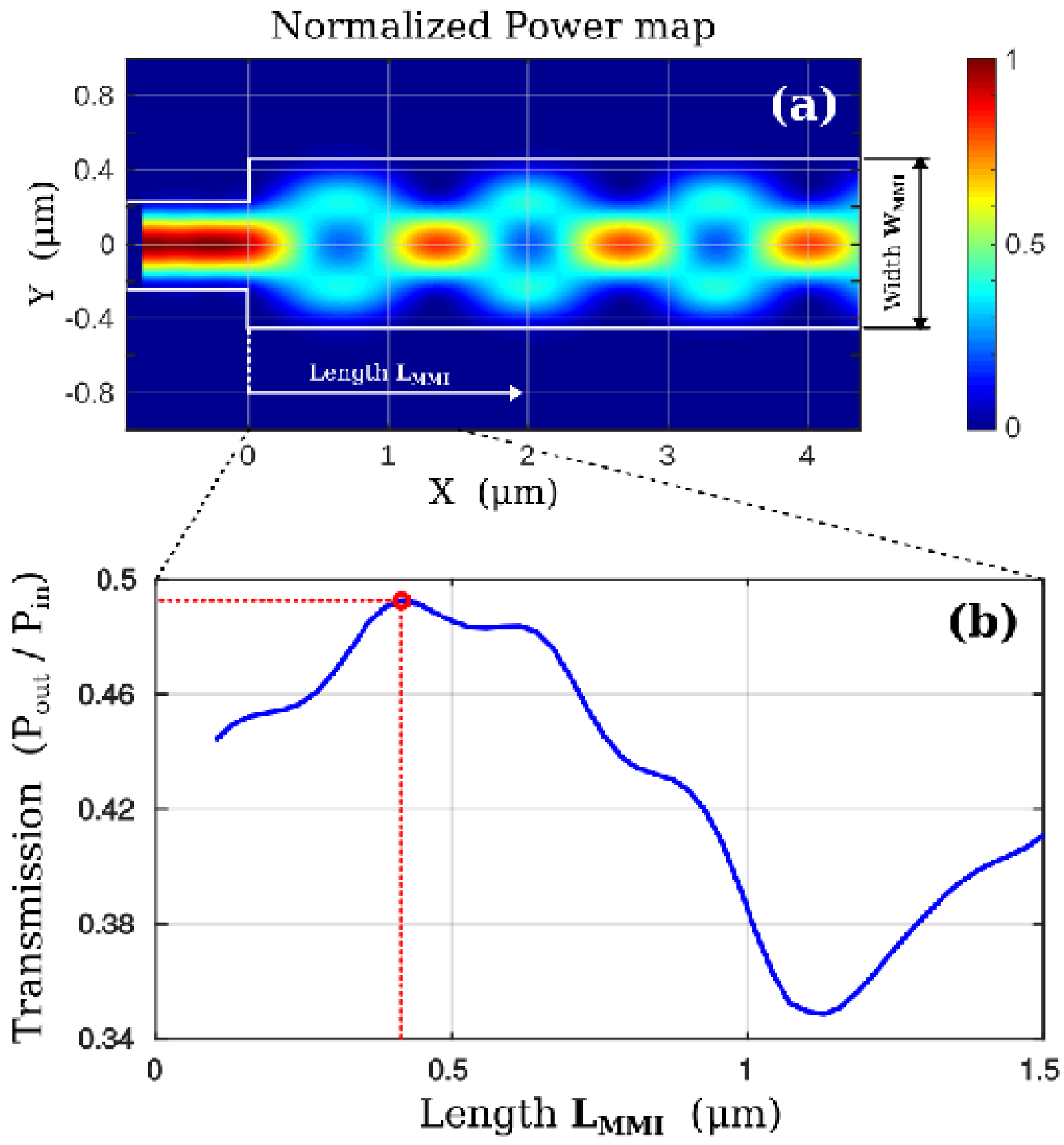

Figure 5 : Tuning the MMI parameters as a function of the propagation distance: Section (a) shows the normalized power map; and panel (b) presents the optical transmission versus MMI´s length (LMMI).

After optimizing all design parameters, we proceed to study the relationship between the transmitted power and the length of the L2 path. This simulation is shown in Figure 6. As can be clearly seen in Figure 6(b), the device presents a periodic transmission that is between -3.05dB and -3.15dB because it prevails the coherent coupling effect. In the right side of Figure 6(a) it is also observed that large L2 segments present peaks in the wavelength transmission, which are not less than -3.22dB. On the other hand, an ultra-compact 1x2N power splitter can be achieved with the proposed and optimized design, for that, it is enough to use N power splitters 1x2 with different L2 segments each one. Finally, the most important feature of this design is the very small footprint; for instance, the footprint of a typical MMI 1x2 with similar behaviour is 6x34.7 µm2, whereas the footprint of the 1x2 power splitter based on the SSC is about 6x4 µm2.

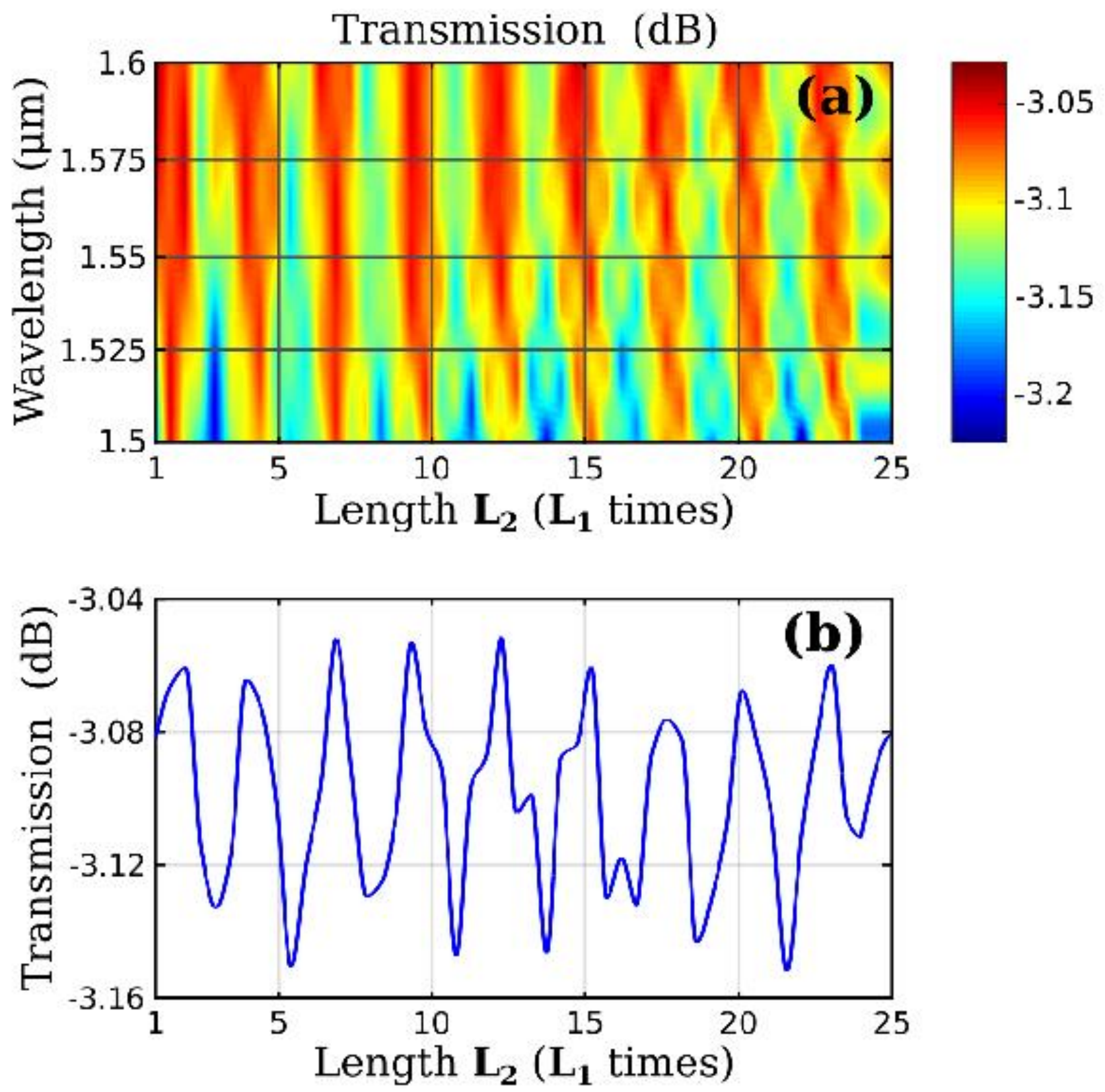

Figure 6: Details of the optical transmission on a branch by sweeping the free parameter L2. This value ranges from 1 to 25 of L1 times. Panel (a) presents a power map with respect to wavelength, while panel (b) shows the total power transmission.

### *3.2 Experimental results*

Considering the above numerical simulations which support a suitable and smaller design for the power splitters analysed in this paper, we fabricated the splitter device follow our proposed power divider system, and for comparison, we also fabricated a standard MMI splitter provided in the manufacturer's library blocks. Based on our simulation results we send to fabricate a photonic chip (10 mm x 15 mm) that includes several systems corresponding to 1x2 and 1x4 splitters for both design architectures, there are SCC and pure MMI power splitters.

Figure 7 shows the measurements corresponding to the optical transmission response (overall losses) for 1x2 splitter for the two models analysed in this work, these are standard MMI and our design SCC. We measured the optical transmission between 1530 nm to 1560 nm for both output ports (A and B) corresponding to each device. In this Figure, the dashed line represents the transmission response obtained from the 1x1 optical circuit used as a reference optical signal for the SOI systems. From this experimental curve we can obtain the relationship between the input/output signals for other fabricated devices to in order to quantify if they were properly designed from our simulations.

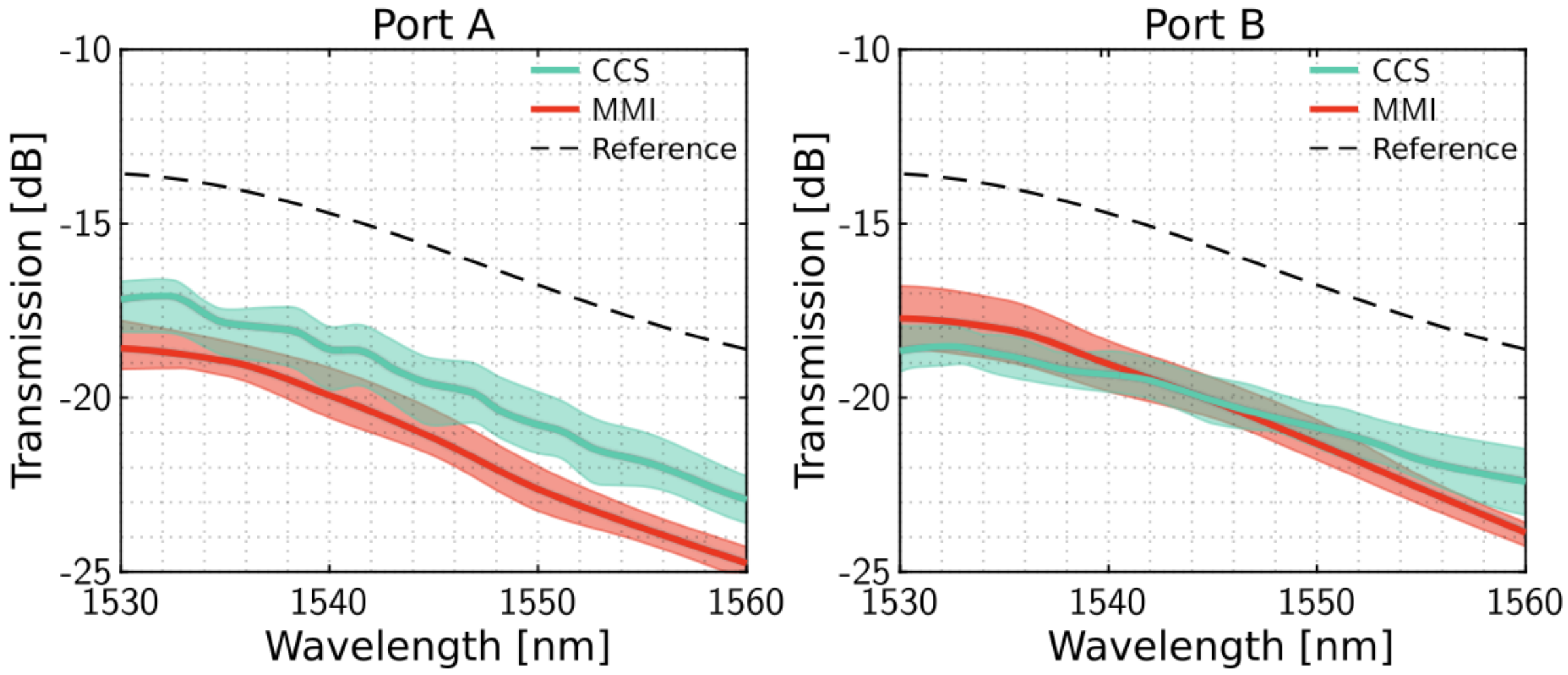

Figure 7. Optical response for the overall transmission versus wavelength, corresponding to port A (left) and B (right) for both the SCC (green) and MMI (red) splitters. The dashed line is the reference optical transmission by the coupling a 1x1 device.

Figure 7 also presents the optical response curve for each port; green curve corresponds to the SCC while red line for MMI device. The solid line and shadow region represents the average signal and the confidence values for each device. As it can be seen, the ports A and B corresponding to the 1x2 splitter fabricated from the two designs conducted in this work, show the same behaviour as function of the input wavelength. However, we can point out that for both devices, port B shows a higher transmission in comparison with port A. This fact could be associated with random distortions because of the technological tolerance for the fabrication process[1].

On the other hand, Figure 8 shows the total optical transmission for each port. In particular, from these resulting curves we have considered the subtraction of the reference transmission curve shown in the dashed line in figure 7 as it was commented. We can observe from the SCC optical response that it is a wavy curve so it could be linked to special geometry used for the optimized design considering SCC effect considering the oscillatory behaviour presented in Figure 3 and Figure 6. This fact is evident in many strong oscillations as a function of the length L1 (L2) parameter for the optical transmission corresponding to the SCC splitter.

As it can be seen, this effect is more noticeable in port A which shows a very noticeable dependence to the propagated wavelength. In contrast, for the standard MMI a clear optical response with wavelength was obtained as it is reported its characteristic curve from many foundries.

1

*Luceda(https://www.lucedaphotonics.com/home).Cornerstone(https://www.cornerstone.sotonfab.co.uk);Intel(https://www.intel.la/content/www/xl/es/architecture-and-technology/silicon-photonics/silicon-photonics-overview.html).

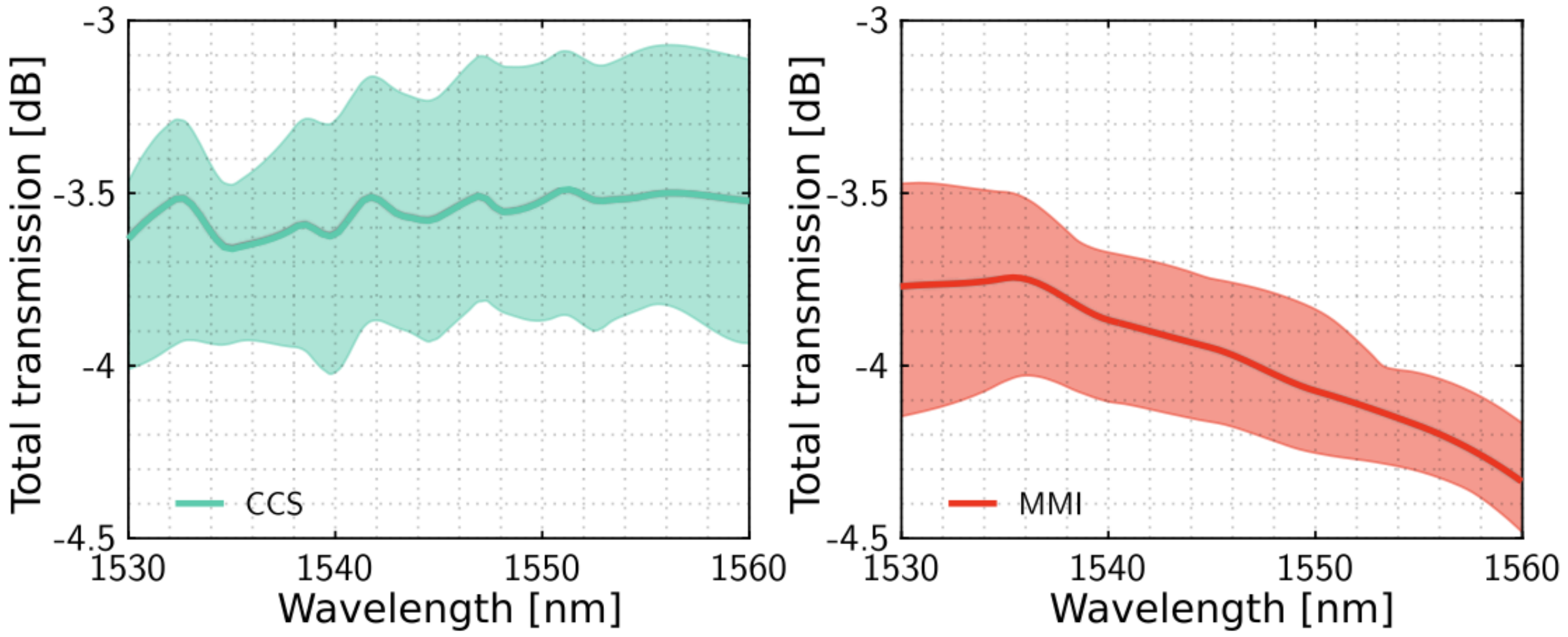

Figure 8. It presents the optical transmission response for every channel versus the propagation wavelength corresponding to CCT (left) and MMI (right).

In order to compare the results obtained from figure 7 and figure 8, we have made a Table which also includes values retrieved from different strategies/technologies form literature to design and fabricate a 1x2 power splitter.

**Table: Optical losses for 1x2 power splitter fabricated by different technologies and/or strategies of design under SOI platform.**

| Design/Technology | optical losses (at 1550 nm) dB/branch | reference |
|---|---|---|
| MMI | 3.06 | [11] |
| MMI | 3.16 | [26] |
| **MMI** | **3.7** | **our work** |
| Y-branch | 3.1 | [25] |
| **MMI+SCC** | **3.6** | **our work** |
| Plasmonic + SOI | 3.1(simulated) | [27] |

Finally, we introduce the results obtained for the 1x4 optical transmission for both photonic designs analysed in this work, which is sketched in Figure 9. For the proposed power splitter: on one side, we can appreciate that all ports show the same behaviour for the optical transmission as a function of the propagation wavelength. However, we can highlight that port B and C shows a slight flatness as a function of the wavelength. The optical transmission spam from -10 to -17 dB (port B and C). In agreement with 1x2 splitter all ports show the higher transmission at the shorter wavelength (1530 nm). For larger wavelengths, the optical transmission drops more than 10 dB (port A and D).

In addition, the 1x4 power splitter made by SCC scheme shows a very low wavy structure observed which could be related to the nature of coherent coupling theory (SCC) for each splitter branch thus it was also shown in Figure 3 and Figure 6 as commented above.

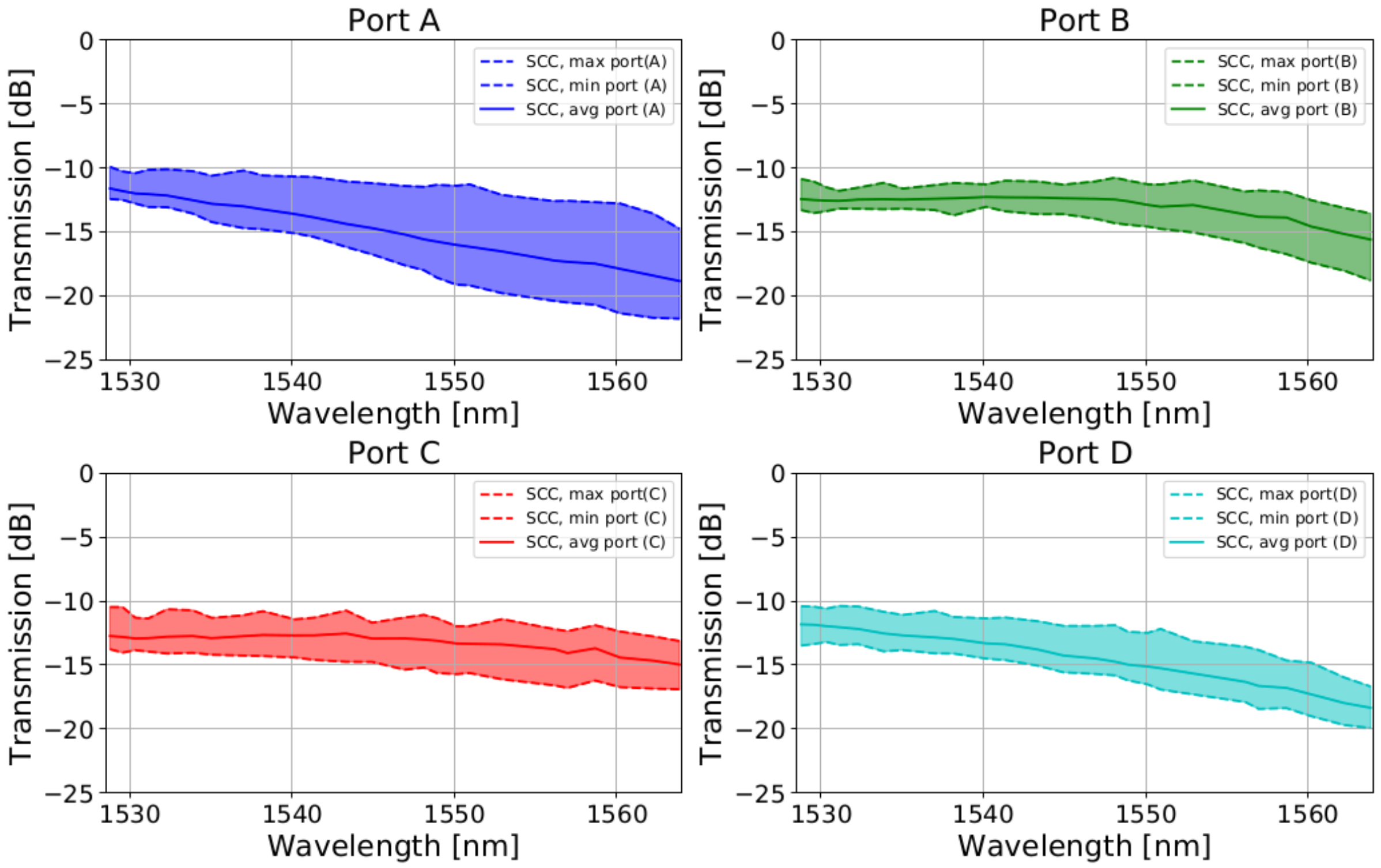

Figure 9. Optical transmission response of SCC 1x4 splitter as a function of the propagated wavelength. Each port was analysed as shown in this figure. The solid line corresponds to the average value while the shadow region represents the confidence range. The upper and lower limits are indicated by dashed lines for each case.

On the other hand, Figure 10 shows the optical response corresponding to the 1x4 MMI splitter. As it can be seen, a similar behaviour than SCC device can be observed. A little flat response can be appreciated from curves corresponding to port B and C. These ports show lower propagation losses since at 1530 nm the optical transmission increases up to 7 dB with respect to wavelength centred at 1560 nm. In addition, port C presents a slight flattening at the transmission as a function of the propagated wavelength in the MMI splitters. This fact can be associated with the random behaviour of the fabrication process and its tolerance in the shape and errors of MMI design which can avoid the usual wavelength response for this type of photonics device.

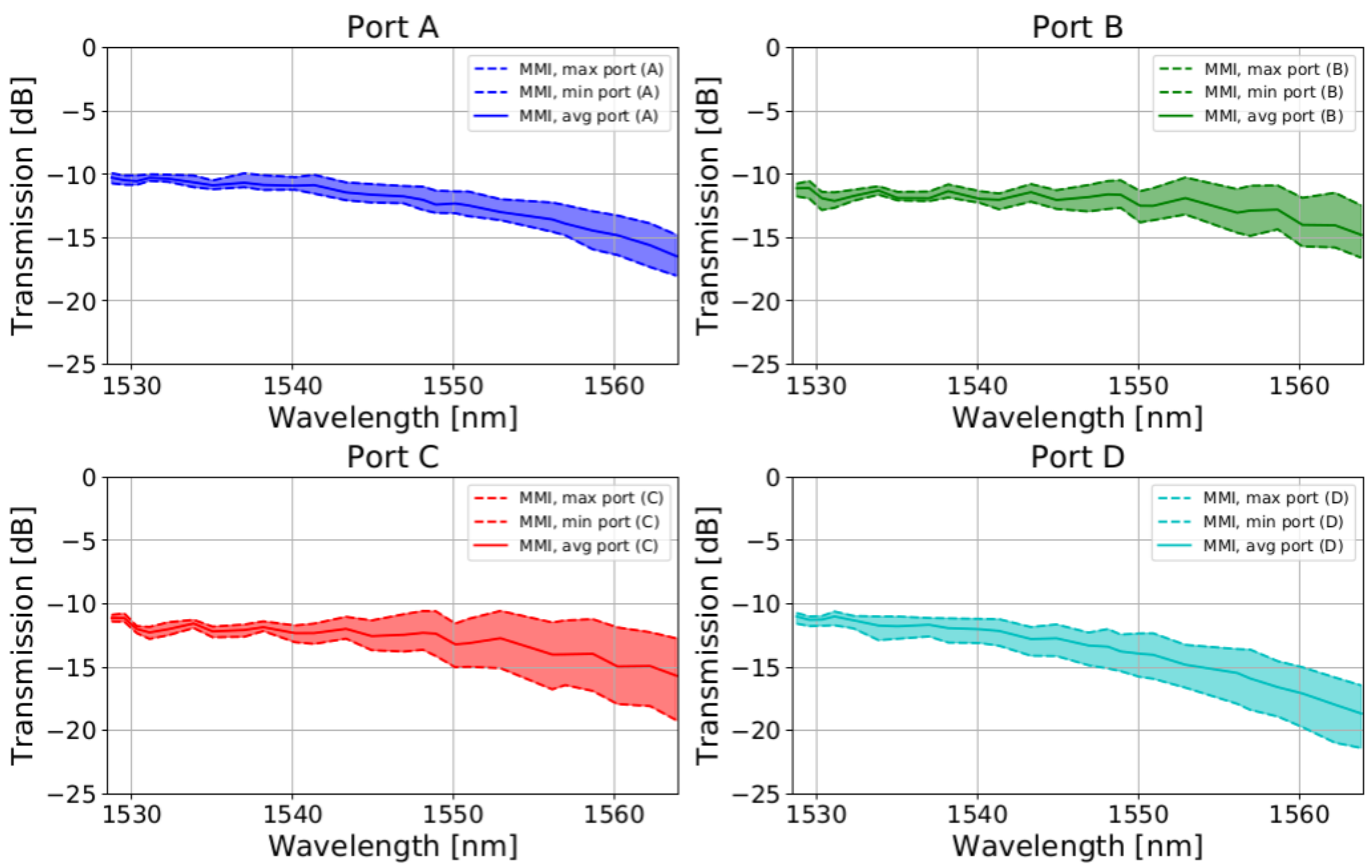

Figure 10. Optical transmission response corresponding to the 1x4 MMI splitter as a function of the propagated wavelength. From each panel, the solid line corresponds to the average data. In addition, dashed upper and bottom lines for each picture represent the confidence range of these experimental data.

## 4- Conclusions:

In this work we have proposed and demonstrated an ultra-compact, nicely adaptable, and highly efficient so-called Y-branch power splitter on silicon-on-insulator. We combine two techniques for that, for one side, the simplified coherent coupling (SCC) to design the output branches. On the other side, we also consider at the entrance a small multimode interference (MMI) structure to split the input light into two guided modes. In order to reduce the footprint and maximize the optical transmission the design parameters were optimized. Besides, a free design parameter to be adjusted according to the application is provided. We have achieved a transmitted power between -3.05dB and -3.15dB, depending on the free design parameter L2.

Complementary, we have fabricated and evaluated the splitter designs, these are: the proposed model of SCC power splitter and compared its performance with those standard MMI splitters achieved in the same photonic chip on SOI platform. In addition, in this paper we compare our splitter architecture and its performance with another integrated signal divider fabricated with several technologies.

In conclusion the innovative design for splitting light results in a very compact system and also shows similar performances that other power splitter models and fabricated by different technologies. It should be remarked that our design proved to be almost nine times smaller than one other commonly used device with similar behaviour.

The proposed strategy of power splitter design opens a new route to design and fabricate integrated photonics circuits under silicon platform, not only by SOI even silicon nitride and silicon oxynitride wafers suitable for exploring other spectral ranges such as mid infrared (MIR) region and enhance the reliability considering the fabrication tolerance for larger photonic devices.

**Acknowledgements.** This work was partially supported by the Agencia Nacional de Promoción Científica y Tecnológica (Argentina) under the project PICT-2016-4086, PICT-2017-0017, and by Universidad Nacional de Quilmes (Argentina) under project PPROF-901-2018 and PROF-827-947/20.